\documentclass[twocolumn,superscriptaddress,showpacs,preprintnumbers,amsmath,amssymb]{revtex4-1}
\usepackage{multirow}
\usepackage[pdftex]{graphicx}
\usepackage{longtable}
\usepackage{dcolumn}
\usepackage{bm}
\usepackage{fancybox}
\usepackage{lineno}

\usepackage{color}

\newcommand{\req}[1]{Eq.(\ref{#1})}
\newcommand{\bfR}{{\bf R}}
\def\connect#1{\leavevmode{\setbox1=\hbox{#1}\copy1%
\raise .2\ht1 \vbox{\moveleft \wd1\vbox{\hrule width \wd1 height .5pt depth 0pt}}%
}}

\begin{document}
\title{Bond-length distributions classified by coordination environments}
\author{Motonari Sawada}
\affiliation{Advanced Mechanical and Electronic System research center (AMES), Faculty of Engineering, Tottori University, Tottori 680-8552, Japan}
\author{Ryoga Iwamoto}
\affiliation{Advanced Mechanical and Electronic System research center (AMES), Faculty of Engineering, Tottori University, Tottori 680-8552, Japan}
\author{Takao Kotani}
\affiliation{Advanced Mechanical and Electronic System research center (AMES), Faculty of Engineering, Tottori University, Tottori 680-8552, Japan}
\affiliation{Center of Spintronics Research Network (CSRN), Graduate School of Engineering Science, Osaka University, Osaka, 560-8531, Japan}
\author{Hirofumi Sakakibara}
\affiliation{Advanced Mechanical and Electronic System research center (AMES), Faculty of Engineering, Tottori University, Tottori 680-8552, Japan}
\date{\today}
\begin{abstract}
We have analyzed bond-length distributions between cations and anions
for crystal structures in the crystallographic open database (www.crystallography.net/cod/).
The distributions are classified by the coordination environments of cations,
which are determined by a tool named as Chemenv (Acta Cryst. (2020). B76, 683-695).
\end{abstract}
\maketitle
\section{introduction}
For finding new useful materials, we have to 
make a plan of experiments and computer simulations 
based on our current knowledge, and execute the plan.
After we renew our knowledge by adding their results,
we again have to make next plan and execute the next plan.
This cycle is repeated until we find new materials.
The purpose of the material informatics (MI) is
to make this cycle as efficient as possible
with minimum human efforts.
We are going to replace what human have to do
with what a computer system can do.

For MI, open database and open-source tools 
become available today. These are going to be integrated as was 
done in pymatgen \cite{ong_python_2013}, ASE \cite{hjorth_larsen_atomic_2017}, and so on.
Thanks to such open resources, we can perform theoretical research to accelerate MI.

The crystal structure database is one of the most fundamental part to support MI.
There are works to analyze crystal structure databases for MI 
\cite{ganose_robocrystallographer_2019,zimmermann_local_2020}.
Recently, we have an open-source database, the crystal open database 
(COD) \cite{vaitkus_validation_2021,quiros_using_2018,
merkys_codcifparser_2016,grazulis_crystallography_2012,grazulis_crystallography_2009}.
It now contains more than four-hundred thousands crystal information files (cifs).
In addition, tools to analyze the crystal structures are available as well.
One of the interesting tool is the Chemenv given by Waronquers et al.
\cite{waroquiers_statistical_2017,waroquiers_chemenv_2020}.
There are tools to specify local environments 
\cite{ganose_robocrystallographer_2019,zimmermann_local_2020}.
Chemenv is one among them to specify the local environment of cations
by the coordination environments (CEs) in the crystal structures.
Chemenv is integrated in pymatgen. 

The local environment of a cation, 
that is, how a cation is surrounded by nearest neighbor (NN) oxygens
is specified by one in a list of CEs. 
The list are illustrated in Figure 1 in Ref.\cite{waroquiers_statistical_2017}.
Specifying a CE gives more detailed description 
about how a cation is surrounded than just specifing the coordination number (CN).
For example, one of CEs, ``T:4 Tetrahedral'', means that
we have four NN oxygens (CN=4) which surround the cation tetrahedrally.

In this paper, as a step along the line of MI, we give a bond-length analysis
classified by CEs which is given by Chemenv.
Our analysis is taken as a extension of the automated analysis of the CEs
given by Waronquers et at \cite{waroquiers_statistical_2017}, 
where they show pie chart each by each for ion species
to show frequencies of CEs in percentage.
We further resolve their pie chart along the axis of bond-length.

Bond-length analysis have been performed by other authors.
One of the latest work is given by Gagn\"e and Hawthorne by a series of papers
\cite{gagne_bond-length_2016,gagne_bond-length_2020}.
With the inorganic crystal structure database (ICSD), they show
the bond-length distribution classified by CN and 
ionic valence given by the bond-valence sum (BVS) method 
with their own parameters \cite{gagne_comprehensive_2015}.
In contrast to their works, we mainly show classified distribution 
just by CEs determined by Chemenv.
Together with our results, we plot the bond lengths determined by Shannon's 
crystal radius
where the bond length is given as the sum of the cation and oxygen crystal radii.

In Sec.\ref{sec:method}, we explain our method.
In Sec.\ref{sec:result}, we show results when cations are alkali, 
alkali-earth, 3d and 4d transition-metal elements in oxides. 
Then we show a discussion for the bond-length distribution classified 
by ionic valence determined by BVS as was done by Gagn\"e and Hawthorne.
We have treated fluorite and nitride in the same manner, 
although we put data of them in the supplemental material \cite{SM}.
In Appendix \ref{anionr}, we show an examination of the crystal radius of anions by
Shannon \cite{Shannon:a12967}.

\section{Method}
\label{sec:method}
We first choose a set of cif files of oxides satisfying following conditions in COD. 
The cif files should be stoichiometric, and be measured at room temperature and 
at ambient pressure; when a cif file contains tags on temperature and on pressure,
the file is picked up only when temperature range is 270 K -- 370 K while
the pressure range is 0 kPa -- 1000 kPa. 
To pick up oxides made from oxygen and cations, we removed files including 
right-side of p-block elements, N, F, P, S, Cl, As, Se, Br, Sb, Te, I, Bi, Po, and At 
in addition to H and C. 
Furthermore, cif files which Chemenv could not recognize is removed, 
probably because of invalid cif files. 
The total number of cif files in the set of oxides is 9265.

In advance to the bond-length analysis, 
we have examined radii of anions as shown in Appendix \ref{anionr}.
Based on the analysis in Appendix, we confirmed that 
the Shannon's crystal radius of oxygen is a reasonable choice.

We apply Chemenv to cif files in the set to get 
bonds between atoms and to determine CEs.
$C_{\rm env}$ is the variable taking
a name in the list of CEs, that is, $C_{\rm env}$ equals one of
\{`S:1 Single neighbor',`L:2 Linear', `A:2 Angular',`TL:3 Trigonal plane',...\};
this list is shown in Figure 1 in Ref.\cite{waroquiers_statistical_2017}. 
At cation site $\bfR$, Chemenv tells us the CE probabilities $P^\bfR(C_{\rm env})$;
for example, it tells us like 
`` $P^\bfR(C_{\rm env})=80.0$\% for $C_{\rm env}=$`L:2' and 
$P^\bfR(C_{\rm env})=20.0$\% for $C_{\rm env}=$`A:2' ''.
The algorithm to determine the probability $P^\bfR(C_{\rm env})$ 
is given in Eq.(1) around in Ref.\cite{waroquiers_statistical_2017}.

With the bond specified by $\bfR i$ where $i$ is the index to distinguish bonds starting from $\bfR$,
we calculate bond weight $w_{\bfR i}(C_{\rm env})$ for a bond $\bfR i$ as
\begin{eqnarray}
w_{\bfR i}(C_{\rm env}) = \frac{P^\bfR(C_{\rm env})}{N_{\bfR} N_{\rm cation}},
\label{eq:weight}
\end{eqnarray}
where $N_{\rm cation}$ is the number of cations in the primitive cell.
$N_{\bfR}$ is the number of bonds starting from the cation $\bfR$. 
Thus the total weight is normalized to be 
$1=\sum_{\bfR i} \sum_{C_{\rm env}} w_{\bfR i}(C_{\rm env})$. 
That is, total weight of bonds from one cif file is unity.
We accumulate $w_{\bfR i}(C_{\rm env})$ 
to the histogram $H_s(l,C_{\rm env})$, where as $l$ is the bond length, $s$ is the species of
cation. We use 0.01 \AA \ for bin width of $l$.
In contrast, Gagn\"e and Hawthorne \cite{gagne_bond-length_2016} use a method to accumulate
a simple weight, unity for a bond, while they accumulate weight classified by CN in addition to by ion valence.
We just use \req{eq:weight} as a natural choice.
We do not claim priority of the weight 
because of difficulty to judge the priority from the view of MI.

\section{Results}
\label{sec:result}
\begin{figure*}[ht]
\includegraphics[width=18cm,height=11cm]{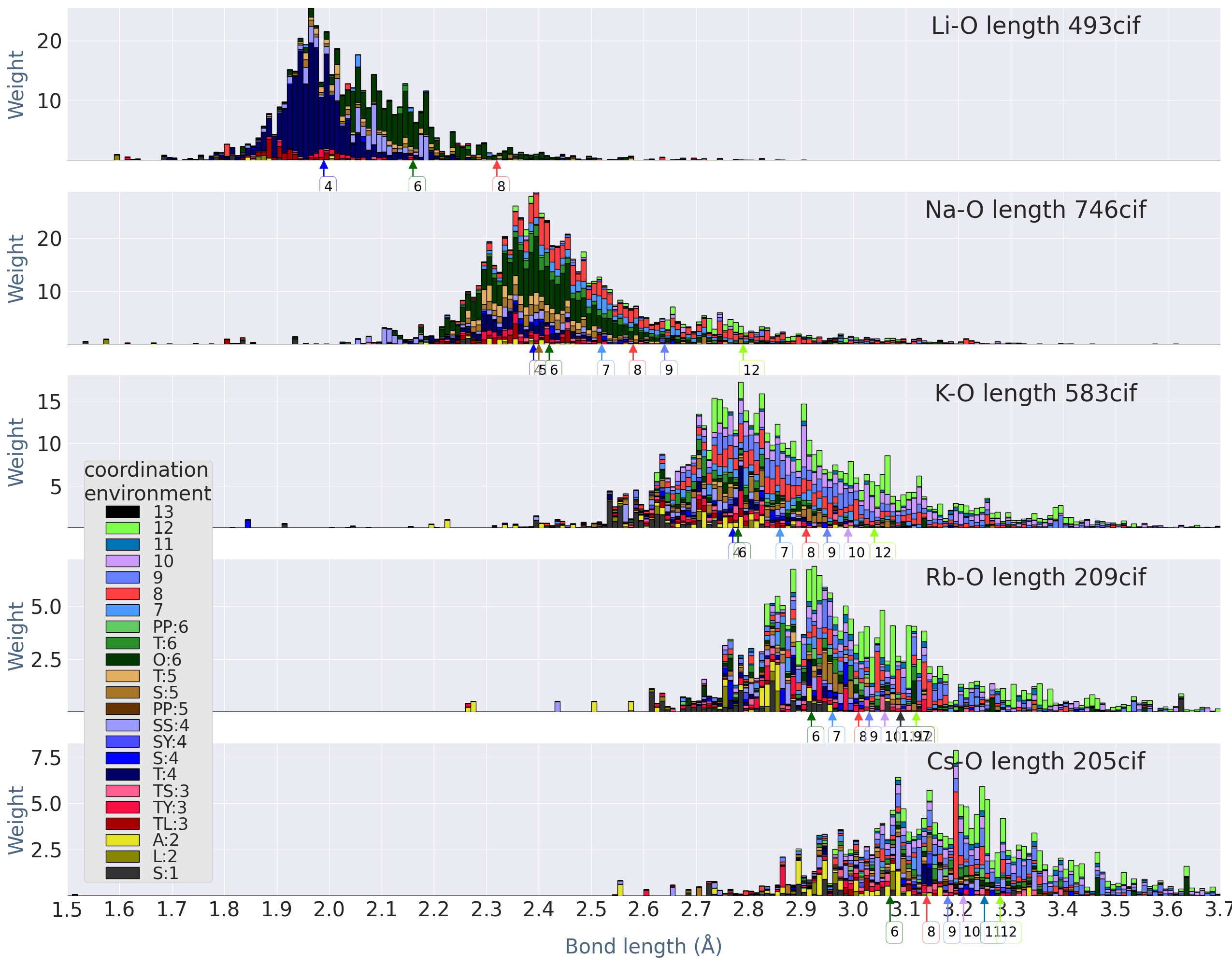}
\includegraphics[width=18cm,height=11cm]{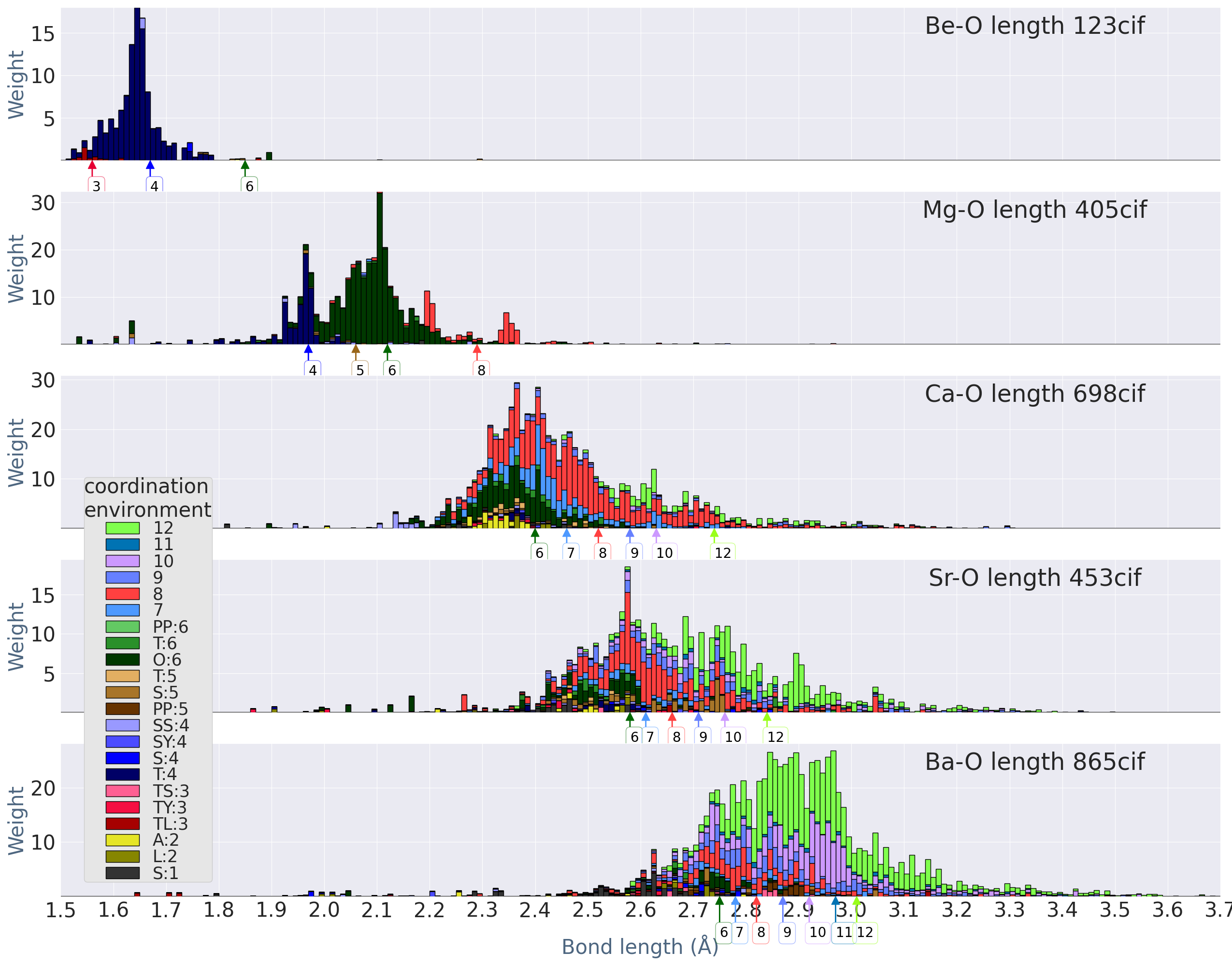}
\caption{\label{fig:LiBe}
Bond length distribution for alkali and alkali-earth elements.
The coordination environments are colored as shown in the inset.
To distinguish similar colors for some CEs, pay attention to the staking orders; 
we use the staking order in the inset.
The bond lengths calculated from Shannon's crystal radius are plotted together, 
specified by arrows whereas numbers in the box show coordination numbers.
}
\end{figure*}

Fig.\ref{fig:LiBe} is for the bond-length distributions 
of alkali and alkali-earth elements.
We show histograms $H_s(l,C_{\rm env})$, where we accumulate weights in \req{eq:weight}.
If we accumulate weight just by CEs neglecting bond-length,
the histograms are in agreement 
with the pie charts in Ref.\cite{waroquiers_statistical_2017},
except the difference due to the difference of used database.
Very outliers shown in our histograms may be or may not be real
because of the validation procedures of cifs in COD 
is still an ongoing problem \cite{vaitkus_validation_2021}.
For example, we actually found cif files of theoretical models,
however, we have not yet removed them because of the difficulty 
to remove them systematically. Thus 
our set of cif files should be refined more in future, however, we believe that
our overall results will not be altered by such refinement 
because such refinements affect only to minor outlines.

We show classification to CEs by stacking histograms in order to have over-all view easily.
Our histograms show chemical trends of bond-length distribution quite well.
We see how the distribution changes for different elements.
For CN$>6$, we did not distinguish CE, just showing CN.
In our histograms, we resolve CN$<$7 into CEs \cite{waroquiers_statistical_2017} 
as
\begin{itemize}
\item S:1 Single neighbor, 
\item L:2 Linear, A:2 Angular
\item
TL:3 Trigonal plane, TY:3 Triangular non-coplanar, TS:3 T-shaped, 
\item
T:4 Tetrahedron, S:4 Square plane, SY:4 Square non-coplanar, SS:4 See-saw, 
\item
PP:5 Pentagonal plane, S:5 Square pyramid, T:5 Trigonal bipyramid,
\item
O:6 Octahedron, T:6 Trigonal prism, PP:6 Pentagonal pyramid,
\end{itemize}
where numbers show CN.

We see some differences from results from Gagn\"e and Hawthorne;
for example, Na-O for Fig.\ref{fig:LiBe} gives a little larger weight for larger bond length 
than Figure 1 in Ref.\cite{gagne_bond-length_2016}, although overall features looks to show
reasonable agreements. 
For example, length for CN=6 are distributed around 2.4 \AA\
in agreement with Figure 5 in Ref.\cite{gagne_bond-length_2016}.
The difference can be because of the difference of weight and how we define bonds.

We can observe chemical trend. 
For Li and Na, we see relatively simple CEs, while K, Rb, Cs 
show variety of CEs.
For Na, we see some amount of weights for both T:5 and S:5.
T:4 is dominant for Li and Be, while Li allows some varieties of CEs such as 
SS:4 and O:6.
The length of Be (CN=4) calculated from Shannon's crystal radii $\sim 1.67$ \AA\ 
shown by the arrow only slightly larger 
than the center of the bond-length distribution. 
The calculated bond length from the radii show similar tendency for other cases.

\begin{figure*}[ht]
\includegraphics[width=18cm,height=11cm]{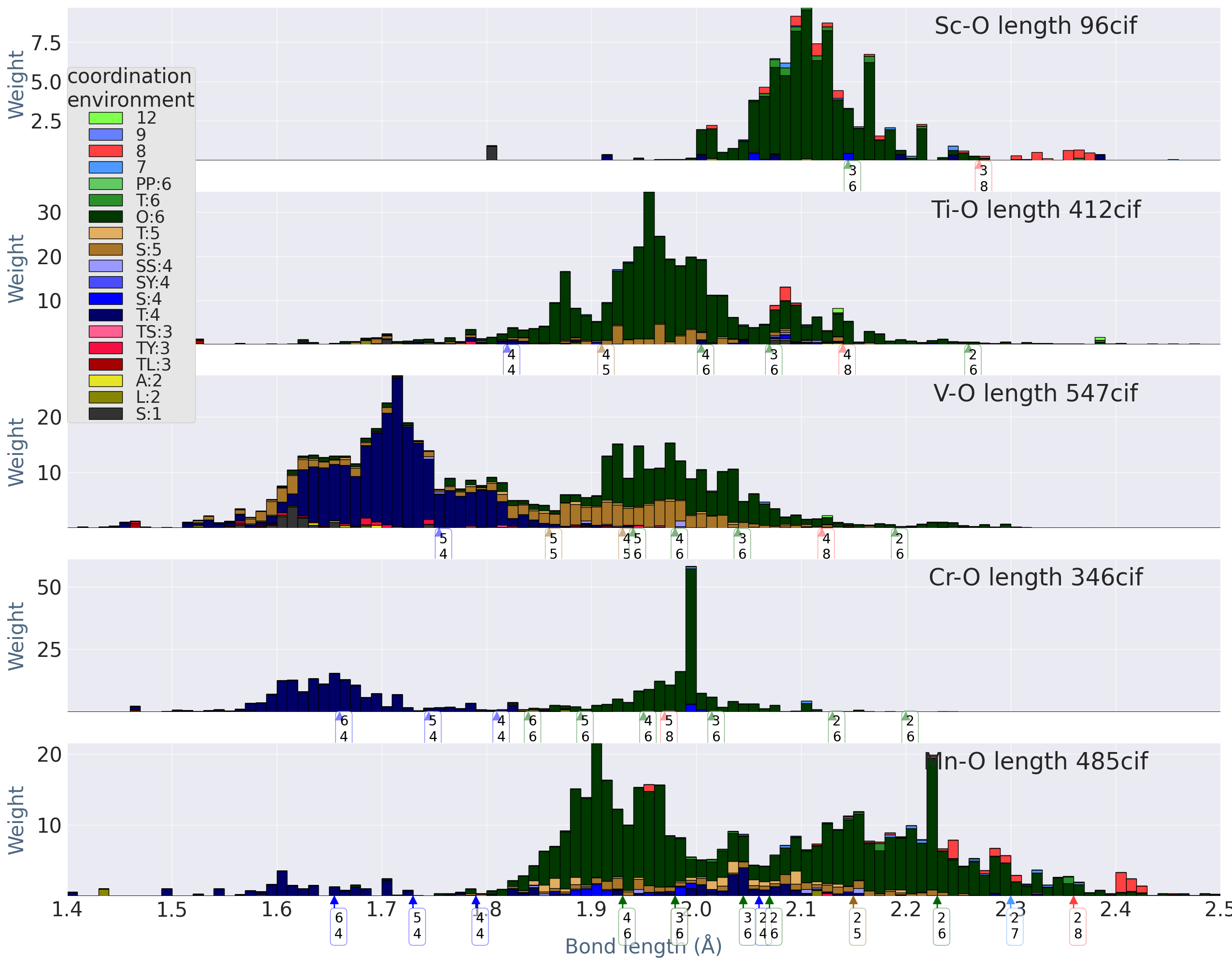}
\includegraphics[width=18cm,height=11cm]{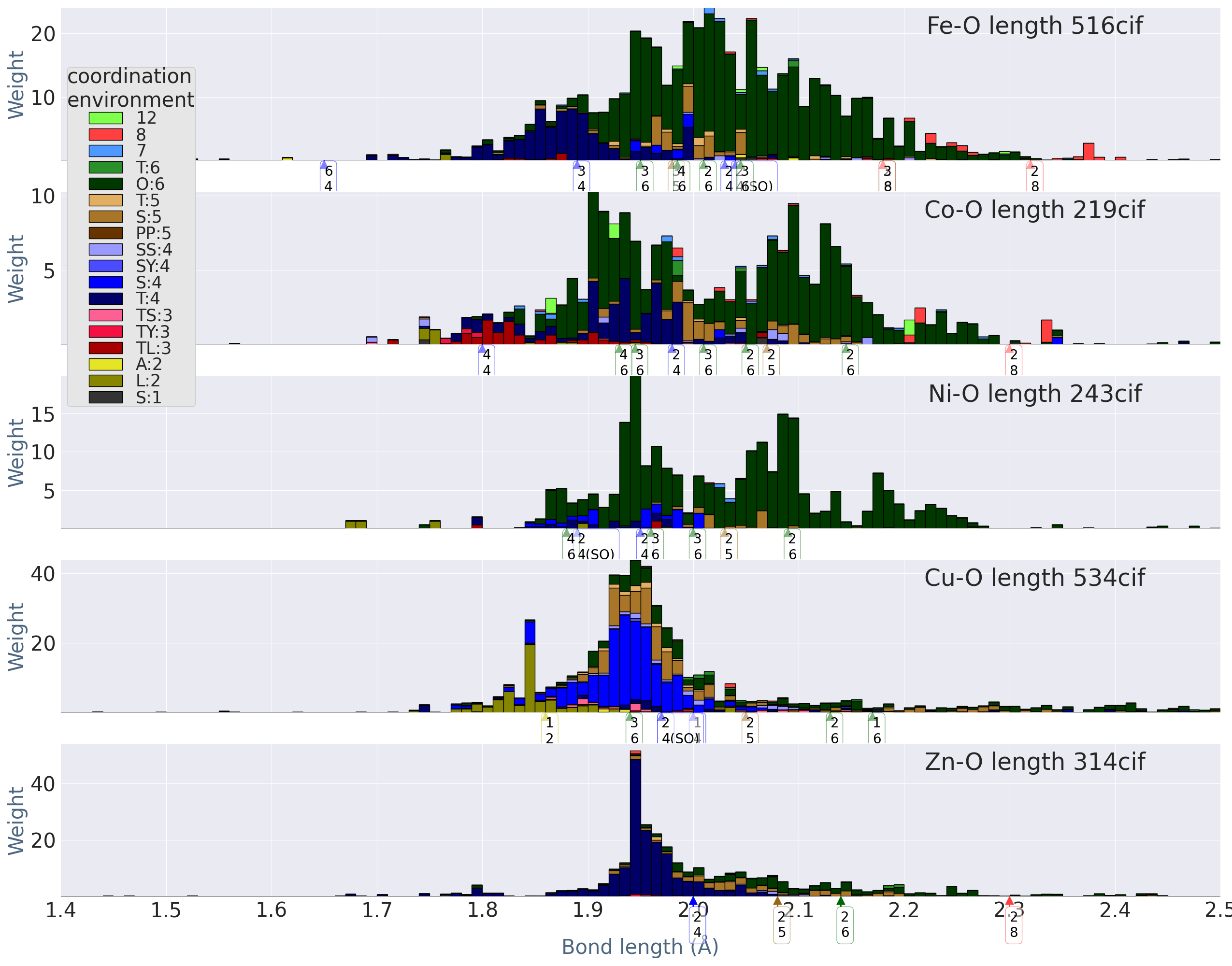}
\caption{\label{fig:ScFe}
Bond length distribution for 3$d$ transition-metal elements. 
See text in Fig.\ref{fig:LiBe}. 
Two numbers below arrows in the box; upper numbers for coordination numbers, lower for ionic valence.}
\end{figure*}

\begin{figure*}[ht]
\includegraphics[width=18cm,height=11cm]{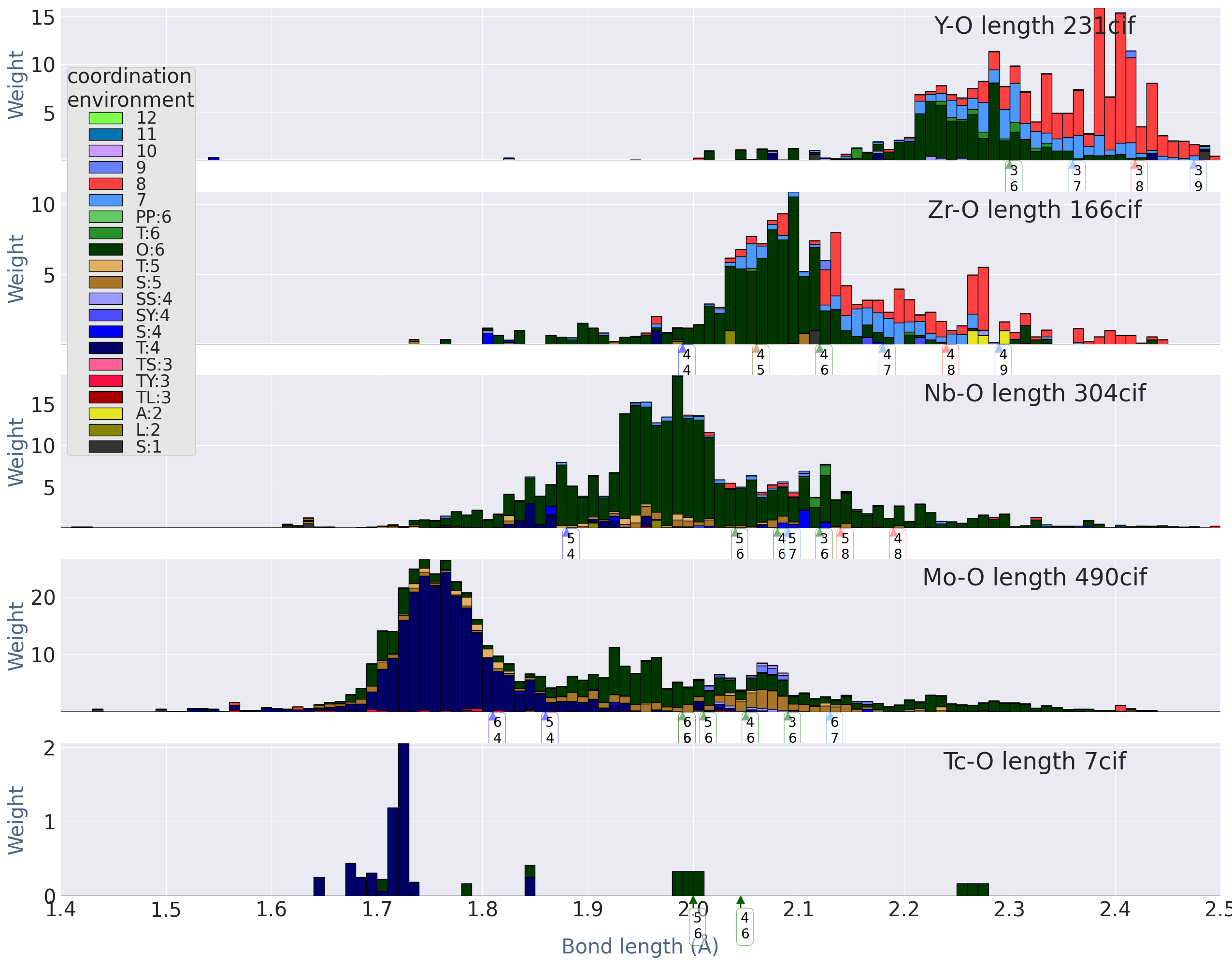}
\includegraphics[width=18cm,height=11cm]{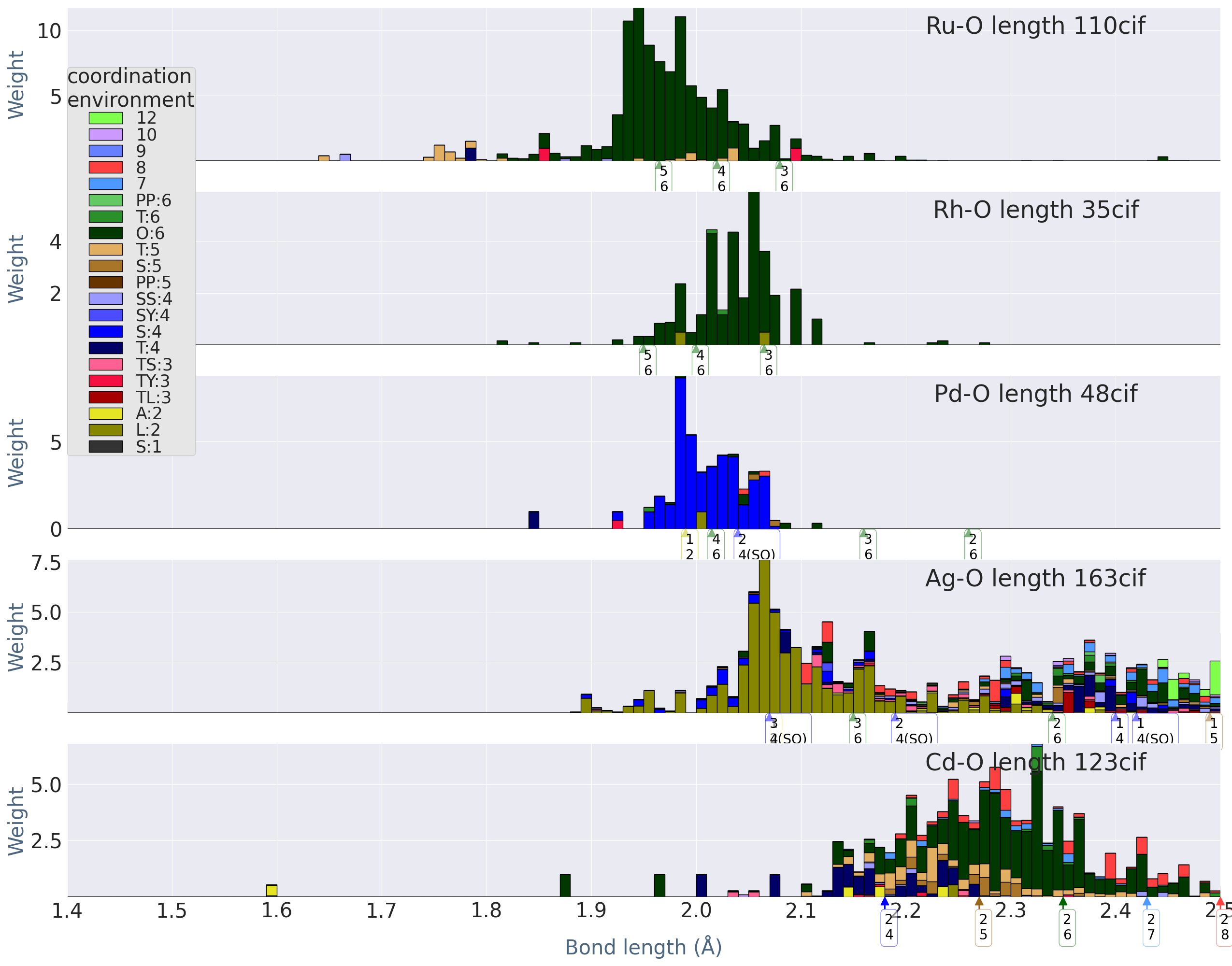}
\caption{\label{fig:YRu}
Bond length distribution for 4$d$ transition-metal elements. 
See text in Fig.\ref{fig:ScFe}. }
\end{figure*}

Figs.\ref{fig:ScFe} and \ref{fig:YRu} are for transition metals.
Our results can be compared with that by Gagn\"e and Hawthorne \cite{gagne_bond-length_2020}.
From the histogram, we can see chemical trends that was not resolved in the pie chart
in Ref.\cite{waroquiers_statistical_2017}. 
The histogram changes as a function of atomic number, somehow systematically.
For example, the peaks of T:4 distribution for V, Cr, and Mn 
are systematically decreasing.
Both Cr and V have peaks of CN=4 and CN=6 at similar lengths, but the widths of 
distributions is quite narrower for Cr. We see some S:5 weights for V.
Histograms for Mn, Fe, Co, and Ni show a systematic changes;
two peaks of O:6 at $\sim$1.9 \AA\ and at $\sim$2.2 \AA\ for Mn may correspond
to two peaks at $\sim$1.95 \AA\ and at $\sim$2.1 \AA\ for Ni.
We see similarity between V and Mo, which belong to the different 
column of periodic table. Ti and Nb show similarity as well. Sc and Zr, too.
From the view of material design, we may like examine 
possibilities of atomic replacement.
For example, the distribution of Ru T:4 and that of Re T:4 is relatively similar,
while allows relatively wide range of length for T:4 for Fe.
This may suggest one of the condition for the replacement 
of these atoms each other at T:4 in crystals.

\begin{figure*}[ht]
\includegraphics[width=5.5cm]{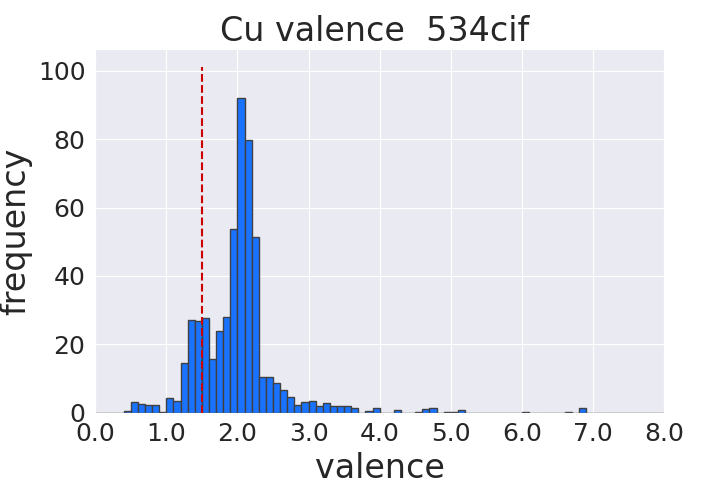}
\includegraphics[width=5.5cm]{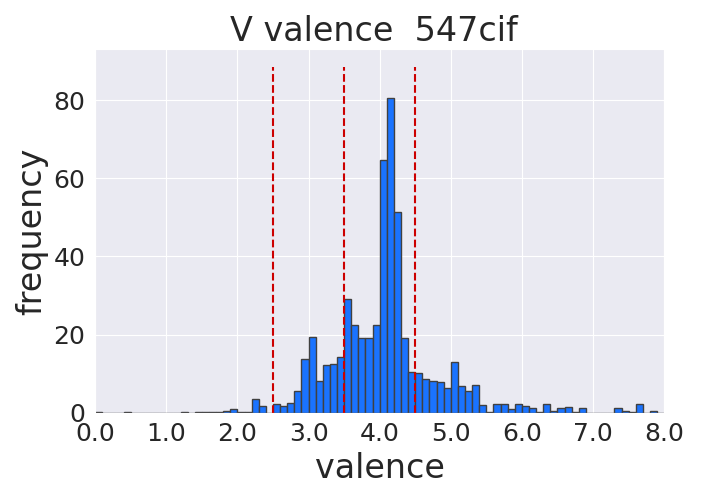}
\includegraphics[width=5.5cm]{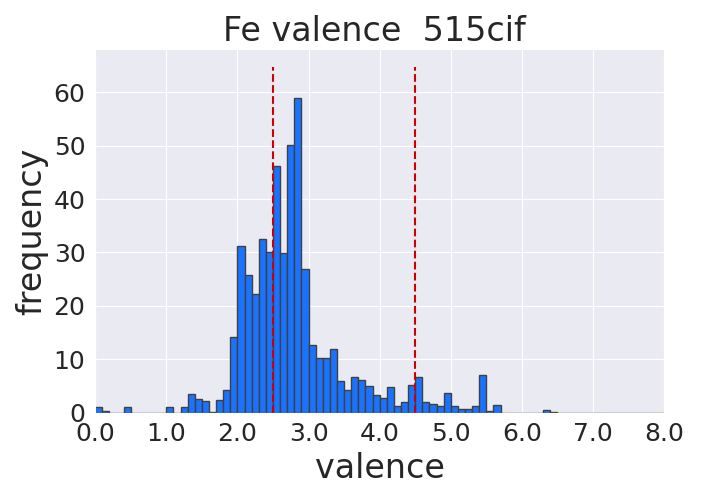}
\includegraphics[width=18cm,height=4cm]{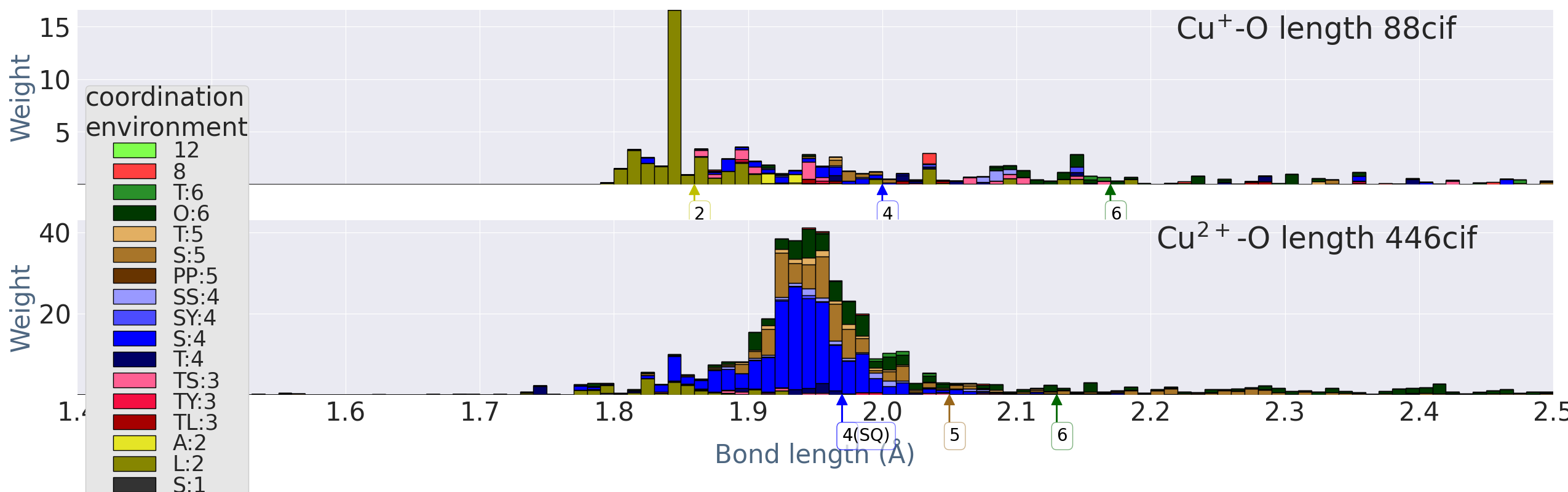}
\includegraphics[width=18cm,height=5cm]{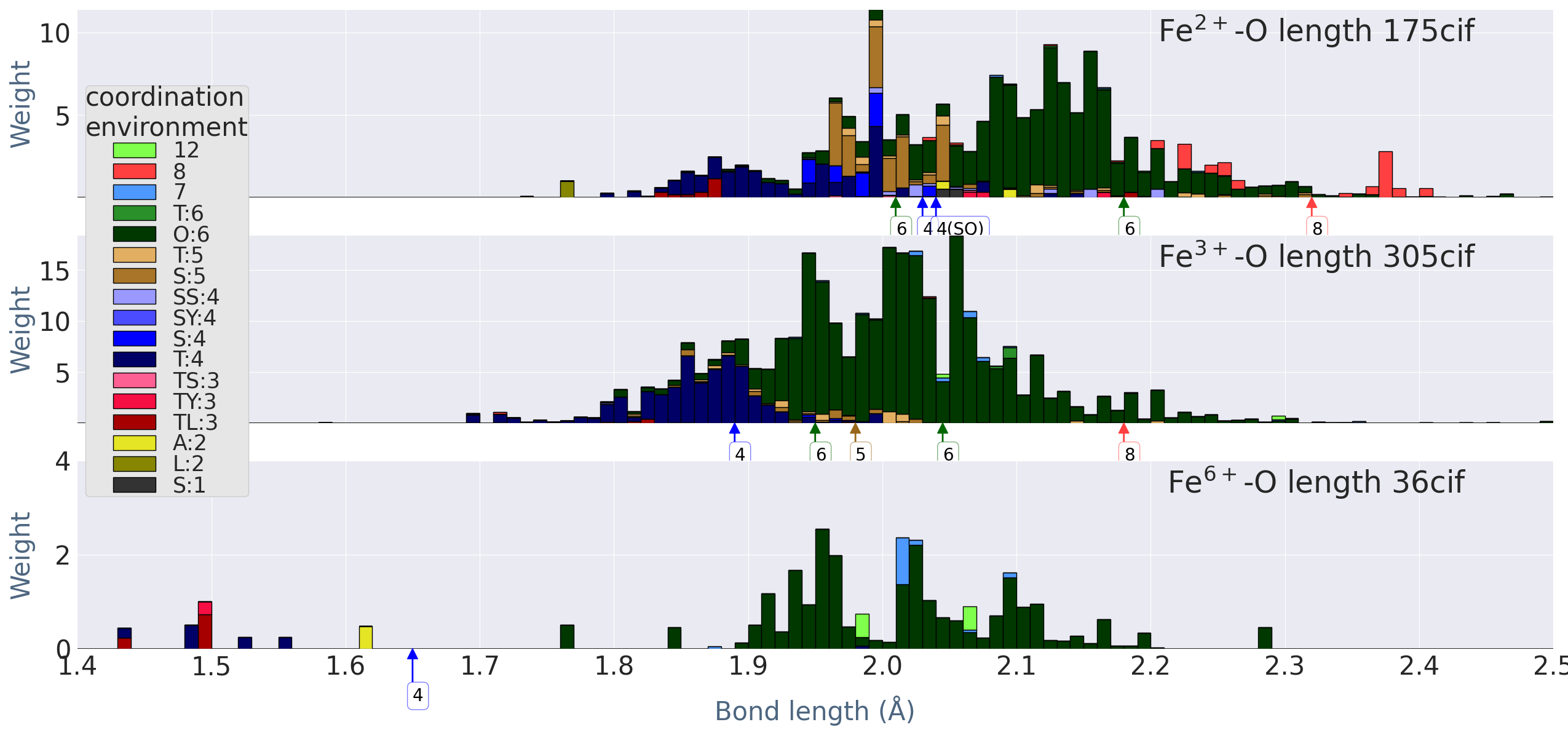}
\includegraphics[width=18cm,height=7cm]{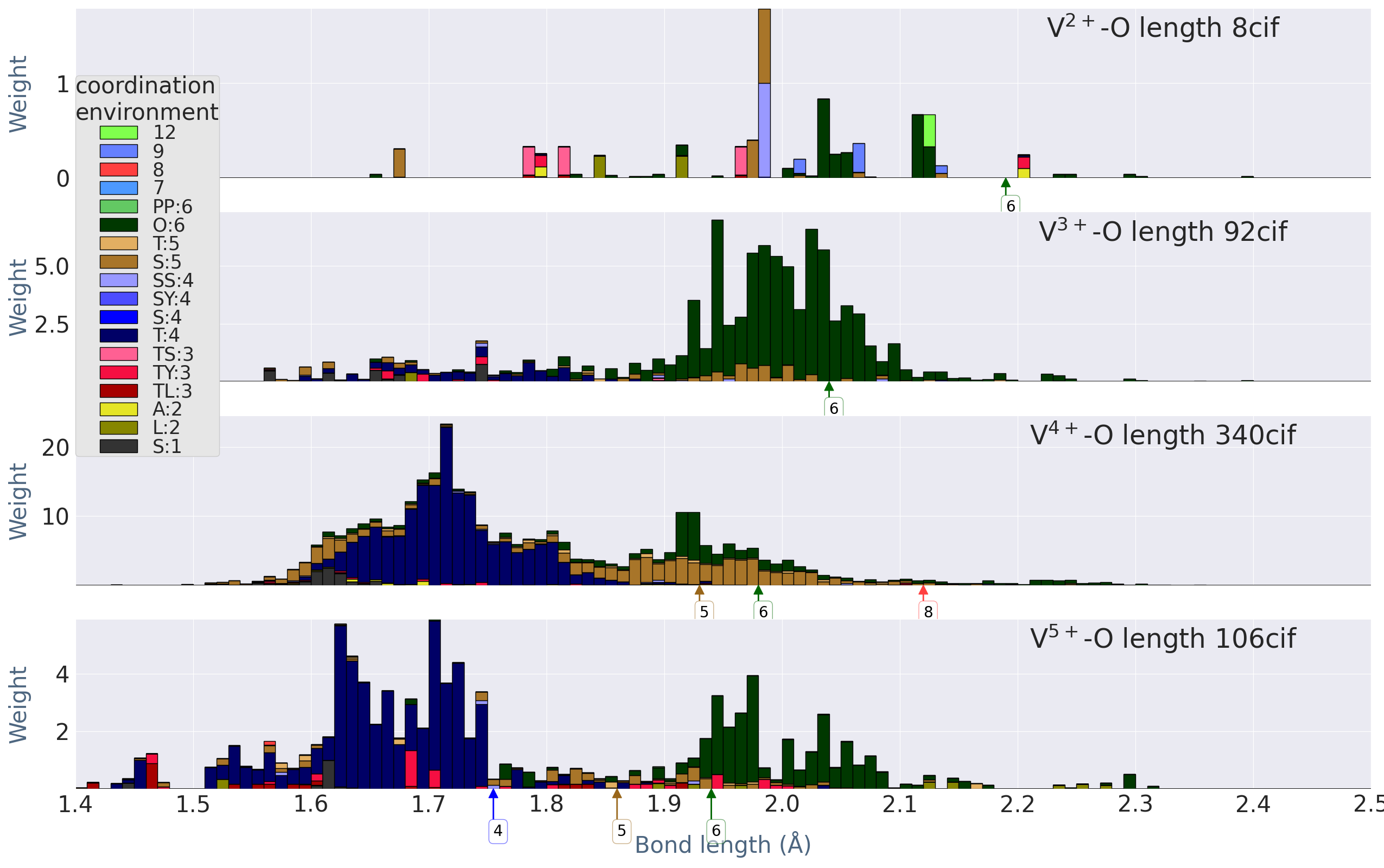}
\caption{\label{fig:CVF}
Bond length distribution of Cu, V, and Fe, resolved by ion valence.
Top three panels show the histogram of BVS; red broken lines are boundaries of
valence for the panels of Cu, V, and Fe.
Sum of the histograms for each elements reproduce Histograms in Fig.\ref{fig:ScFe}.
}
\end{figure*}

Fig.\ref{fig:CVF} is further classification 
of V, Fe and Cu of Fig.\ref{fig:ScFe} into ion valences,
which are determined by the BVS whose parameters are determined in Ref. 
\cite{gagne_comprehensive_2015}. 
This is in order to evaluate the idea of BVS. 
Note that BVS is based on a simple idea to determine the charge 
transfer via a bond between atoms as a function of length.
BVS differentiate Cu$^+$ and Cu$^{2+}$ well. 
Based on the formula of BVS, 
smaller valence means longer bond length and/or smaller CN.
Thus it is reasonable that CN=2 is mainly selected in the case of Cu$^+$. 
We see that shorter bond length are selected for larger valences for Fe and V.

The valence given by BVS is just a parameter to measure how a cation is surrounded.
Longer bond length and less number of neighbors gives the same effects to BVS.
For example, we can keep the same BVS of a cation 
when we move four neighbor oxygens away a cation
but move another two neighbor oxygens closer simultaneously; 
BVS claims that such a move do not change the environment 
of cation in the sense of valence. This sounds reasonable qualitatively, 
however, we should note the limitation on BVS.

Our histograms have the information of bond length and CEs, without classified by BVS.
When we have a new material, its CEs and bond lengths are
compared with other materials so as to judge similarity and differences.
Our histograms should be a help to perform such comparison.


\section{Summary}
We have given the histograms of the bond-length distributions of COD,
classified by CEs.
Our histograms show systematic and non-systematic 
features of the histograms dependent on species of cations.
Our analysis should be a clue to use the bond-weight distribution
as the finger print of materials.
From the opposite point of view, our challenge is to predict crystal structures 
as accurate as possible for given composition formula.
We believe such ability of prediction can be enhanced 
with further analysis of crystal structure database along the line
of our analysis.

\ \\
We thank to Prof. Masami Kanzaki for helpful discussions, especially for BVS.
This work was partly supported by JSPS KAKENHI Grant Number 17K05499.

\appendix
\section{An analysis of anion radius}\label{anionr}
We have performed an analysis to confirm Shannon's anion radius.
If we assume rigid-sphere model that anions are close packed whereas cations are small enough, 
we can calculate anion ion radius $R$ from the cell volume $V_{\rm cell}$
and number of anion atoms per cell $N$ from
\begin{eqnarray}
0.740 V_{\rm cell} = \frac{4 \pi R^3 N}{3} .
\label{eq:orad}
\end{eqnarray}
Here 0.740 is the ratio of the close-packed rigid spheres. 

\begin{figure}[ht]
\includegraphics[width=8.5cm,height=3cm]{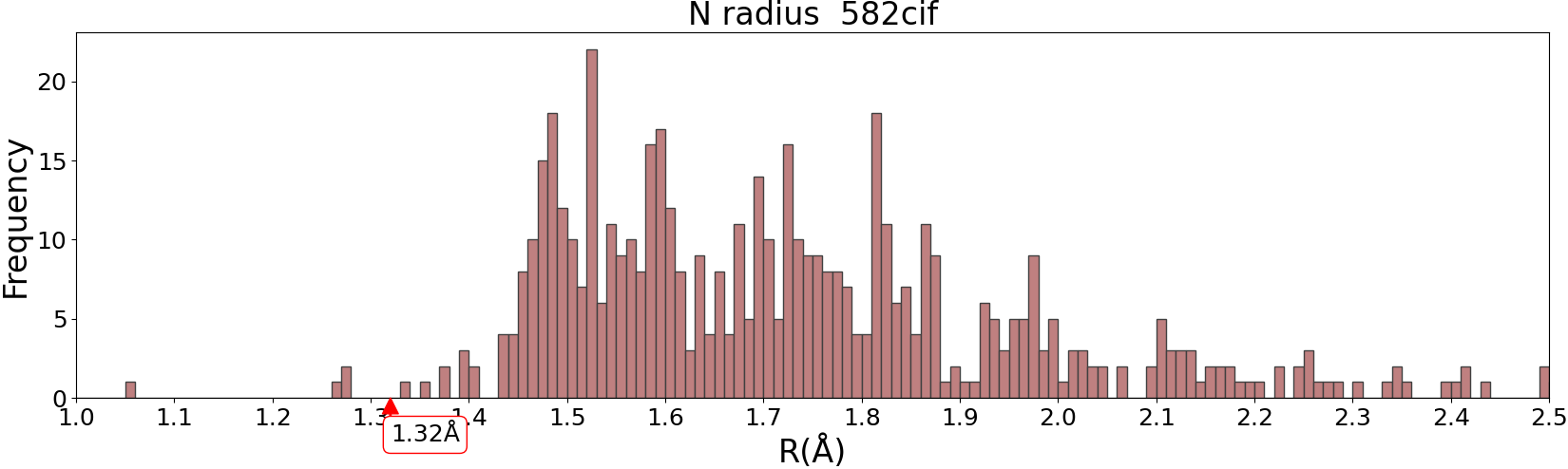}
\includegraphics[width=8.5cm,height=3cm]{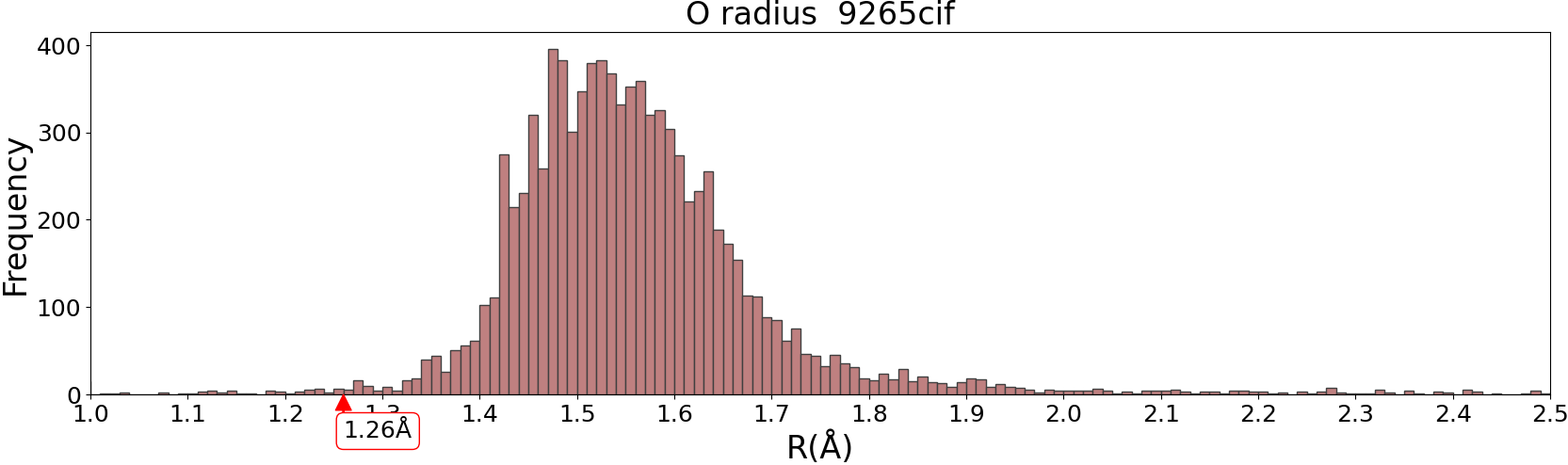}
\includegraphics[width=8.5cm,height=3cm]{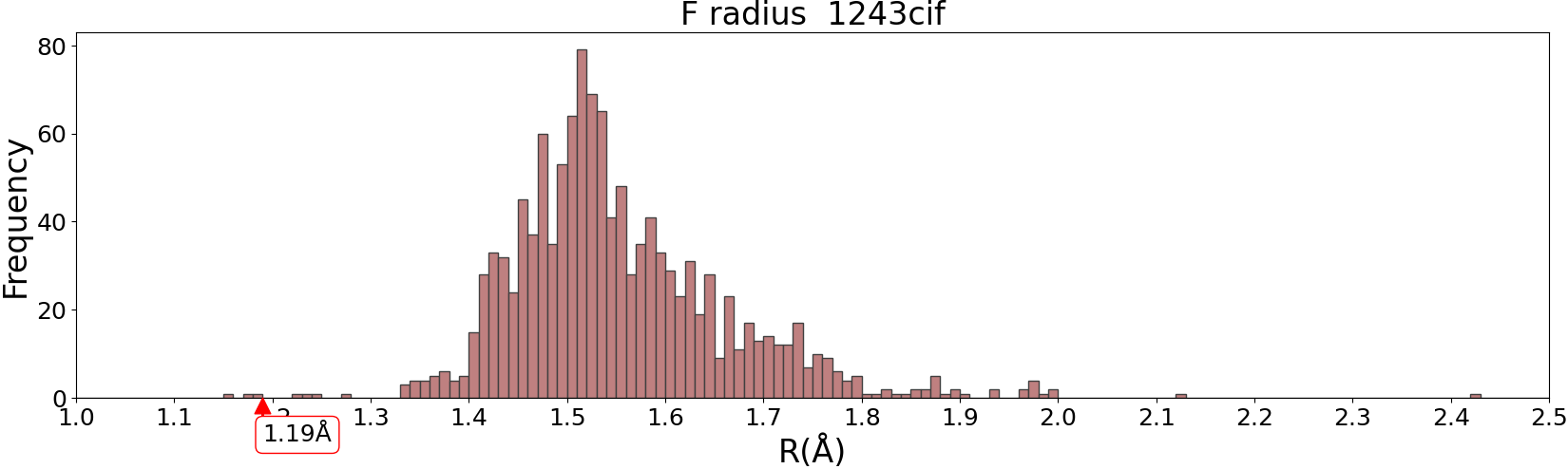}
\caption{\label{fig:anionr}
We plot anion radius $R$ in \req{eq:orad}, together with Shannon's crystal radius
\cite{Shannon:a12967,Shannon2}.
Nitrides and Fluorite, as well.}
\end{figure}

Fig.\ref{fig:anionr} show file frequency histogram as functions of $R$. 
For oxides, we see that $R=1.26$ \AA\ is located almost at the right end of the tail of the distribution,
while it shows sudden jump up just around $R \sim 1.4$ \AA. 
$R=1.26$ \AA\ is the crystal radius while $R=1.4$ \AA\ is the ionic radius
given by Shannon \cite{Shannon:a12967,Shannon2}. 
The sudden jump up may correspond to the idea of rigid (or very hard) 
sphere packing, thus we may like to use 1.4 \AA\ as the radius of oxygen ion.
However, if we use oxygen-ion radius,
cif files classified below $R=1.4$ \AA (more than 400 cif files) are already over-packed. 
To avoid this problem, the crystal radius $R=1.26$ \AA\ should be a better choice.
Fig.\ref{fig:anionr} may indicate that anion spheres are not hard enough after all as was assumed
in the rigid sphere packing model.
For nitrides and fluorite, we see that crystal radii are small enough as in the case of oxygen.

\bibliography{refbond}
\end{document}